\documentclass[12pt]{article}

\usepackage[T1]{fontenc}
\usepackage[utf8]{inputenc}
\usepackage[english]{babel}
\usepackage[font=small, labelfont = bf]{caption}
\usepackage{graphicx}
\usepackage[table]{xcolor}

\usepackage {graphicx,color}
\usepackage{capt-of}
\usepackage{booktabs}
\usepackage{varwidth}

\usepackage{amssymb}
\usepackage{amsmath}
\usepackage{amsbsy}
\usepackage{setspace}
\usepackage[numbers]{natbib}
\usepackage{bm}
\usepackage{textcomp}
\usepackage{amsthm}
\usepackage{hyperref}
\usepackage{floatrow}
\newfloatcommand{capbtabbox}{table}[][\FBwidth]
\usepackage{blindtext}

\numberwithin{equation}{section}
\newtheorem{thm}{Theorem}[section]
\newtheorem{lemma}{Lemma}[section]
\newtheorem{condition}{Condition}[section]
\newtheorem{proposition}{Proposition}[section]
\newtheorem{corol}{Corollary}[section]
\newtheorem{remark}{Remark}
\newtheorem{algorithm}{Algorithm}[section]

\newtheorem{example}{Example}

\newtheorem{definition}{Definition}

\newcommand{\beq}{\begin{equation}}
\newcommand{\eeq}{\end{equation}}
\newcommand{\beas}{\begin{eqnarray*}}
\newcommand{\eeas}{\end{eqnarray*}}
\newcommand{\bea}{\begin{eqnarray}}
\newcommand{\eea}{\end{eqnarray}}
\newcommand{\bei}{\begin{itemize}}
\newcommand{\eei}{\end{itemize}}
\newcommand{\ben}{\begin{enumerate}}
\newcommand{\een}{\end{enumerate}}
\newcommand{\bet}{\begin{theorem}}
\newcommand{\eet}{\end{theorem}}
\newcommand{\bel}{\begin{lemma}}
\newcommand{\eel}{\end{lemma}}
\newcommand{\bep}{\begin{proposition}}
\newcommand{\eep}{\end{proposition}}
\newcommand{\bed}{\begin{definition}}
\newcommand{\eed}{\end{definition}}
\newcommand{\bec}{\begin{corollary}}
\newcommand{\eec}{\end{corollary}}
\newcommand{\bex}{\begin{example}}
\newcommand{\eex}{\end{example}}

\newcommand{\argmax}{\mathop{\rm arg\max}}

\begin{document}

\title{Testing and estimation for clustered signals }
\author{Hongyuan Cao$^1$ and Wei Biao Wu$^2$}

\footnotetext[1]{~ Department of Statistics, Florida State University, Tallahassee, FL 32306.}
\footnotetext[2]{~ Department of Statistics, University of Chicago, Chicago, IL, 60637.}
%
\date{}
\maketitle
%
%
%
%
%

\begin{abstract}
We propose a change-point detection method for large scale multiple testing problems with data having clustered signals. Unlike the classic change-point setup, the signals can vary in size within a cluster. The clustering structure on the signals enables us to effectively delineate the boundaries between signal and non-signal segments. New test statistics are proposed for observations from one and/or multiple realizations. Their asymptotic distributions are derived. We also study the associated variance estimation problem. We allow the variances to be heteroscedastic in the multiple realization case, which substantially expands the applicability of the proposed method. Simulation studies demonstrate that the proposed approach has a favorable performance. Our procedure is applied to {an array based Comparative Genomic Hybridization (aCGH)} dataset. 
\end{abstract}


 \thispagestyle{empty}

%
%
%
%
%
\vspace{-1cm}
\section{Introduction}\label{intro.sec}
\vspace{-0.5cm}
\setcounter{page}{0}
Signal detection and multiple testing in a data rich environment have been important research topics in natural and social sciences. Typical examples include detecting anomalous traffic in computer networks \cite{szor05}, identifying voxels that correlate with certain activities \citep{heller06} in functional Magnetic Resonance Imaging (fMRI), and associating single nucleotide polymorphisms (SNPs) with clinical outcomes \citep{pritchard01}. The predominant framework in these research is via individual analysis--testing each hypothesis separately and declaring statistical significance if the $p$-value is less than certain threshold \citep{bh95} or the two-sample $t$-statistic falls into the rejection region \citep{fan07, ck11}. Various approaches were proposed to improve the power by incorporating structured or prior information. For example \cite{caisun09} studied group hypothesis testing; \cite{genovese06, hu10} investigated $p$-value weighting; \cite{caowu15} considered $p$-value aggregating; \cite{duzhang14,lirina15} utilized prior experimental information on each hypothesis in the inference stage with data from a new experiment; and \cite{liuandshao2014, liu15} harnessed the sparsity of mean vectors with student's $t$-statistics. 

For data with signals having clustered structure, multiple testing approaches currently in use fall into two general classes. The first approach defines possible regions of interest in advance, either by field knowledge or an independent experiment. \cite{sun15} proposed a spatial testing procedure with pre-specified regions of interest in a compound theoretical framework; \cite{heller06} developed an algorithm specifically tailored for brain imaging data where a preliminary scan is used to select clusters by grouping highly correlated voxels; \cite{perone04} used the supreme statistic in a random field to construct confidence envelopes for the proportion of false discoveries and \cite{benjamini07} used a two-stage hierarchical testing procedure to test predefined clusters first followed by a trimming stage to clean locations in which the signal is absent. 
The second approach is to adaptively identify a collection of differentially behaved regions with proven false discovery rate control. For example \cite{shen02} mapped the data in the wavelet domain first and removed redundant hypotheses to reduce the number of hypotheses tested and improve power; \cite{zhangcm10} studied multiple testing via false discovery rate control for large scale imaging data; \cite{siegmund11} treated each cluster as a testing unit and defined the false discovery as the clusters that are falsely declared among all declared clusters under the assumption that the number of false discoveries is approximately Poisson. 
The Poisson approximation requires the sparsity assumption on the signals. \cite{chouldechova14} gave a summary of literature in this area and developed new tools for spatial multiple testing. In this line of research, a cluster is defined to be a true discovery if it has non-zero overlap with the support of the signal. Methods that try to incorporate cluster size to improve power were also explored in \cite{chouldechova14}.

In this paper, we study multiple testing problems for data with clustered signals. We propose a new test statistic that adaptively recognizes such clusters. Our test statistic aggregates information along a sliding window to boost signal noise ratio. 
At the boundary between signal and non-signal segments, the test statistic can be much larger than it is within the non-signal cluster. We investigate the asymptotic distribution of the proposed test statistic and set up rejection criterion controlling certain type I errors. A new algorithm is proposed to locate signal clusters for followup studies. We do not require signals to be sparse, which may be especially valuable given the current conjecture of polygenic effects on complex disease \citep{zhou13}. Furthermore, we allow signals to vary within a cluster, which differs from the popular assumption that the means are identical within the same cluster \cite{yao1993, yao1988}. Numerical studies show that when signals have varied sizes, the proposed method has increased detection accuracy compared to method that assumes same signal size within a cluster \cite{yao1993, wuandzhou2018, wuandzhou2020}. Computationally, the speed of our algorithm is linear with number of tests while the algorithm used in \cite{yao1993} is quadratic. Unlike \cite{caowu15}, we present a new approach for variance estimation under the setup of multiple testing. This is accomplished through the order statistics of the average squares of the original data across a sliding window, which is consistent under certain regularity conditions for the one realization case. In addition, we consider the multiple realization scenario and allow the variance to be heteroscedastic. New test statistics are proposed with unknown parameters consistently estimated with available data to conduct statistical inference. Moreover, the newly proposed algorithms are more accurate in detecting break-points than algorithms proposed in \cite{caowu15} as an additional turning parameter $\delta$ is used in the maximization to locate break-points. Numerical studies show improved detection precision compared to methods that did not utilize the clustering structure \cite{fan07}. 

%


%
Recent multiple testing procedures that incorporate covariates require estimation of the prior probability that the $i$th test corresponds to a null, $i =1, \ldots, m.$ These weights are then estimated adaptively from available data. In particular, \cite{Tansey2018} uses an empirical Bayesian two group mixture model and proposes to minimize a penalized likelihood function where fused lasso type of penalty is used to have spatial smoothing \citep{tibshiraniwang2008}. OrderShapeEM proposed by \citep{cao2021} imposes a monotone increasing constraint on the prior probability of being null and a monotone decreasing constraint on the density function of $p$-value under alternative distribution.  The implementation is achieved through combination of EM algorithm and pool-adjacent-violator-algorithm (PAVA). AdaPT \citep{leiandfithian2018} requires an order of the $p$-value to incorporate external information to boost power. SABHA \citep{liandbarber2019} modifies the BH procedure by incorporating the probability that the $i$th test corresponds to a null, $i = 1, \ldots, m.$  SABHA further suggests different ways to estimate such probabilities, including ordering, grouping, and low total variation. AdaPT and SABHA achieves finite sample control of FDR.  

In our work, we impose block signal structure to improve power. Different from covariate adjusted multiple testing, we do not use covariate for individual test, instead, we treat clustered signals through aggregation of individual $p$-values. We do not require external covariate, such as ordering. Our results are asymptotic in terms of number of test $m$. 

An important method for spatial cluster detection is based on scan statistics \citep{patil03, szor05}. Scan statistic is defined as the maximum number of points in a fixed window as the window is shifted across the domain. The $p$-value is computed under the uniform distribution on the domain and the threshold is designed to control the familywise type I error. This statistic is used for an omnibus test of the null hypothesis that there is no clustering. If the test rejects the null hypothesis, then it leaves open the question of where and how much clustering exists. Our test statistic is devised to compare the observed information with its expected value under the null hypothesis that there is no signal and then take the maximum across the domain. If the omnibus test detects signals, our proposed algorithm can locate such signals which is of special interest for followup studies.

{The rest of the paper is organized as follows. In Section \ref{sec:onereal}, we introduce a structured hypothesis testing problem with one realization. Section \ref{sec:multisep} studies the case that there are multiple realizations. In Section \ref{sec:03161012}, we examine the performance of the proposed procedure via simulation; we see that our procedure is better able to detect clustered signals and the variance estimate has a good performance. Section \ref{sec:real} presents an application of the methodology to an array based Comparative Genomic Hybridization (aCGH) dataset. }


%

\vspace{-1cm}
\section{Test and estimation with one realization}
\label{sec:onereal}
\vspace{-0.5cm}
In this section we shall first present a structured hypothesis testing problem with locally clustered signals. Suppose we are given noisy data of the form
\begin{equation}\label{model}
X_j = \mu_j + Z_j, \, 1\le j \le p,
\end{equation} 
where $Z_j$ are i.i.d. with mean $0$ and variance $\sigma^2$, and $\mu_j$ are means or signals. We say that a signal is present at location $j$ if $\mu_j \not= 0$. In the study of aCGH data, we let $X_j$ be the $\log_2$ ratio between the test and the reference sample intensities at locus $j$. Then $X_j > 0$ (resp. $X_j < 0$) means copy number duplication (resp. deletion). In this section we assume that one realization $(X_j)_{j=1}^p$ is available. In Section \ref{sec:multisep} we shall deal with the situation that multiple realizations are available with possibly non-i.i.d. $Z_j$. Based on the observation $(X_j)_{j=1}^p$, we test the null hypothesis of no signal 
\begin{eqnarray}
\label{eq:0416031306}
H_0: \, \mu_1 = \cdots = \mu_p = 0 
\end{eqnarray}
versus the alternative hypothesis that signals are clustered: there exist {\it break-points} $1 = \tau_0 \le \tau_1<\cdots <\tau_l \le \tau_{l+1} = p$ such that
\begin{eqnarray}
\label{eq:alt}
H_1: \mu_1 = \cdots = \mu_{\tau_1-1}=0,& & \mu_{\tau_1}, \cdots, \mu_{\tau_2-1}\ne 0, \cr
\mu_{\tau_2} = \cdots =\mu_{\tau_3-1}=0, & & \mu_{\tau_3}, \cdots, \mu_{\tau_4-1} \ne 0, \cdots.
\end{eqnarray}
Let ${\cal S}_f = \{\tau_f, \ldots, \tau_{f+1} - 1\}$, $f = 1, 2, \ldots$. We call sets ${\cal S}_1, {\cal S}_3, \ldots$, {\it signal clusters} on which $\mu_j$s are non-zero and let ${\cal S} = {\cal S}_1 \cup {\cal S}_3 \cup \ldots$ be the signal set. Let ${\cal N} = {\cal S}_0 \cup {\cal S}_2 \cup \ldots$ be the non-signal set. Note that our definition of break-points is different from change-points that are used in change-point analysis, where the alternative hypothesis is typically formulated as
\begin{eqnarray*}
\label{eq:altcpa}
H_c: \mu_1 = \cdots = \mu_{\tau_1-1} \not= \mu_{\tau_1} = \cdots = \mu_{\tau_2-1}\not =
\mu_{\tau_2} = \cdots =\mu_{\tau_3-1} \not=\cdots.
\end{eqnarray*}
For example, if there exists a $j$ in the signal cluster ${\cal S}_1 = \{\tau_1, \ldots, \tau_2 - 1\}$ of (\ref{eq:alt}) such that $\mu_{\tau_1} = \ldots = \mu_j \not= \mu_{j+1} = \ldots = \mu_{\tau_2-1}$, then this $j$ is a change-point while it is not a break-point in our sense. While providing a very general framework, our setting of allowing unequal $\mu_j$s in the signal clusters substantially complicates the related statistical inference. The primary goal of the paper is to test $H_0$ vs $H_1$ and to locate those break-points. 

%
%

\vspace{-0.5cm}
\subsection{One-sided Test}
\label{sec:11}
If in the signal sets ${\cal S}_1, {\cal S}_3, \ldots$, all non-zero $\mu_i$ are positive, namely 
\begin{eqnarray}\label{eq:0416031319}
H'_1: \mu_1 = \cdots = \mu_{\tau_1-1}=0,& & \mu_{\tau_1}, \cdots, \mu_{\tau_2-1}> 0, \cr
\mu_{\tau_2} = \cdots =\mu_{\tau_3-1}=0, & & \mu_{\tau_3}, \cdots, \mu_{\tau_4-1} > 0, \cdots,
\end{eqnarray}
then we can use the following test statistic
\begin{eqnarray}
\label{eq:0416031321}
R_{i}^\circ = \frac{1}{k}\sum_{j=i+1}^{i+k} X_j,
\end{eqnarray}
{where $k$ is the window size parameter}. Note that the mean \textcolor{red}{is} $E R_{i}^\circ = k^{-1} \sum_{j=i+1}^{i+k} \mu_j$. Intuitively, $i$ can be classified in the signal cluster if $R_{i}^\circ$ is big. The cutoff values can be computed based on Theorem \ref{th:0416032233}, which provides a uniform Gaussian approximation of the distribution of $R_{i}^\circ$. Theorem \ref{th:0416032233} follows from Theorem \ref{th:KC} with $n = 1$. For completeness we state it here. It asserts that under $H_0$, $R_{i}^\circ$ can be uniformly approximated by the Gaussian process
\begin{eqnarray}
\label{eq:0416031337}
\sigma G^\circ_i = \frac{1}{k}\sum_{j=i+1}^{i+k} \sigma \eta_j, \mbox{ where } 
 \eta_j \overset{i.i.d.}{\sim} N(0, 1).
\end{eqnarray}
We shall quantify the closeness by the coupling distance
\begin{eqnarray}
\label{eq:0416071927}
\Delta^\circ = \sqrt k \max_{0 \le j \le p-k} |R_{j}^\circ / \sigma - G^\circ_j| 
\end{eqnarray}
and the distributional distance
\begin{equation}
\label{eq:0416071928}
\rho^\circ = \sup_{u} |P(\sqrt k \max_{0 \le j \le p-k} R_{j}^\circ / \sigma  \ge u)
 - P(\sqrt k \max_{0 \le j \le p-k} G^\circ_j \ge u) |.
\end{equation}

We first introduce a moment condition.

\begin{condition}
\label{cond:0416032220}
$Z_1, Z_2, \ldots$, are i.i.d. with mean $0$ and variance $\sigma^2$, and the $\theta$th norm $\| Z_i \|_\theta := (E |Z_i|^\theta )^{1/\theta} < \infty$, where $\theta > 2$. Write $K_\theta := \| Z_i \|_\theta$. 
\end{condition}

\begin{thm}
\label{th:0416032233}
Assume Condition \ref{cond:0416032220} and $\mu_i = 0, 1\le i \le p.$ (i) Let $\theta > 2$. Then there exists a possibly larger probability space on which one can define $(Z_j)_j$ and $(\eta_j)_j$ such that, for all $u > 0$ {and any positive integer $k$,}
\begin{eqnarray}
\label{eq:0416032236}
P\left[ k^{1/2} \Delta^\circ \ge c_0 u \right]
 \le { \frac{p K_\theta^\theta}  {u^\theta \sigma^\theta} },
\end{eqnarray}
where $c_0$ is a constant only depending on $\theta$. (ii) Let $\theta > 3$. The distributional distance
\begin{equation}\label{eq:0416032235}
\rho^\circ \lesssim 
 { k^{-1/6} (\log p)^{7/6} 
 + (p k^{-\theta/2})^{1/(\theta+1)} (\log p)^{(3 \theta-2) /(2+2\theta)} }
\end{equation}
where {$a \lesssim b$ means a = O(b)} and the multiplicative constant in $\lesssim$ only depends on $\theta$, $\sigma^2$ and $K_\theta$. Namely there exists a constant $C > 0$ depending on $\theta$, $\sigma^2$ and $K_\theta$ such that $\rho^\circ \le C ( k^{-1/6} (\log p)^{7/6} + (p k^{-\theta/2})^{1/(\theta+1)} (\log p)^{(3 \theta-2)/(2+2\theta)}).$
\end{thm}

Theorem \ref{th:0416032233} implies that, if the window size $k$ satisfies $p^{2/\theta} = o(k)$, then $\Delta^\circ = o_P(1)$ by letting $u = (p k)^{1/(\theta+2)}$. Under the slightly stronger condition
\begin{eqnarray}
\label{eq:0416082200}
p^{2/\theta} (\log p)^{3-2/\theta} = o(k),
\end{eqnarray}
we have $\rho^\circ = o(1)$, suggesting that $R_{j}^\circ$ and $\sigma G^\circ_j$ are uniformly close.

Let $\hat \sigma^2$ be an estimate of $\sigma^2$ and $g_{1-\alpha}$ be the $(1-\alpha)$th quantile of $\max_{0 \le j \le p-k} G_j^\circ$, $\alpha \in (0, 1)$. The latter can be computed by Monte Carlo simulations. In Section \ref{sec:varest} we shall propose a consistent estimate of $\sigma^2$ when $(\mu_j)_j$ has form (\ref{eq:alt}). Theorem \ref{th:0416032233} suggests rejecting $H_0$ and accepting the alternative hypothesis $H'_1$ of (\ref{eq:0416031319}) at level $\alpha$ if
$\max_{0 \le j \le p-k} R^\circ_j > \hat \sigma g_{1-\alpha}$.
Alternatively, let $T = p / k$, by Corollary A1 in \cite{br73} we can also have the Gumbel convergence
\begin{equation}
\label{eq:0416032249}
P[{ {\max_{0 \le j \le p-k} \sqrt k G^\circ_j } \over \sqrt{2 \log T} }
 - 1 - { {\log \log T - {1\over 2} \log (4 \pi)} \over {4 \log T}} \le v ]
\to e^{-e^{-v}},
\end{equation}
which gives an approximate solution for $g_{1-\alpha}$ by letting $v = - \log\log (1-\alpha)^{-1}$. We do not recommend the latter since the convergence of (\ref{eq:0416032249}) is very slow. A bootstrap calibration procedure is proposed in Section \ref{sec:cutoff} which has better finite sample properties.


\subsection{Two-sided Test}
\label{sec:12}
Under the general alternative $H_1$ of (\ref{eq:alt}), the test statistic (\ref{eq:0416031321}) is no longer applicable since the $\mu_j$s in the signal clusters can potentially cancel each other out. As a simple remedy, assuming at the outset that $\sigma^2$ is known, we define the modified version  
\begin{eqnarray}
\label{eq:0316302218}
R_i^\dagger = \frac{1}{k}\sum_{j=i+1}^{i+k} (X_j^2 - \sigma^2),
\end{eqnarray}
which, since $\epsilon_j = Z_j^2 - \sigma^2 + 2 \mu_j Z_j$ has mean $0$ under $H_0$, mimics $R_i^\circ$ in (\ref{eq:0416031321}) in view of
\begin{eqnarray}
\label{eq:0416032319}
X_j^2 - \sigma^2 = \mu_j^2 + (Z_j^2 - \sigma^2 + 2 \mu_j Z_j) = \mu_j^2 + \epsilon_j.
\end{eqnarray}
Hence a location $i$ with a big value of $R_i^\dagger$ will likely be in signal clusters, regardless of signs of $\mu_j$. Then we can apply the one-sided test procedure in Section \ref{sec:11}. Note that the other modified version $R_i^\star := k^{-1}\sum_{j=i+1}^{i+k} (|X_j| - m_1)$, where $m_1 = E |Z_j|$, does not have the property that $E( |X_j| - m_1) = E (|\mu_j + Z_j| - |Z_j|) > 0  $ for non-zero $\mu_j$. So in general $R_i^\star$ cannot be used in the two-sided test. Assume that $E(Z^4_j) < \infty$. Similar to (\ref{eq:0416071927}) and (\ref{eq:0416071928}), we define
\begin{eqnarray}
\label{eq:kappa}
\Delta^\dagger = \sqrt k \max_{0 \le j \le p-k} |R_{j}^\dagger / \kappa - G^\circ_j| , \mbox{ where } \kappa = \|Z_j^2 - \sigma^2 \|_2 = [E(Z_j^2 - \sigma^2)^2]^{1/2}, 
\end{eqnarray}
and the distributional distance
\begin{eqnarray*}
\rho^{\dagger} = \sup_{u} |P(\sqrt{k} \max_{0 \le j \le p-k} R_j^{\dagger}/\kappa \ge u) - P(\sqrt{k} \max_{0\le j \le p-k} G_j^{\dagger} \ge u)|.
\end{eqnarray*}
Note that under $\mu_j=0, $ we have $\mbox{Var}(\epsilon_j) = E(Z_j^2 - \sigma^2 + 2\mu_jZ_j)^2 = \kappa.$
\begin{corol}
\label{th:0416032326}
Assume Condition \ref{cond:0416032220} hold with $\theta > 4$ and $\mu_i = 0, 1\le i \le p.$ Then there exists a larger probability space on which one can define $(Z_j)_j$ and $(\eta_j)_j$ such that for all $u > 0$,
\begin{eqnarray}
\label{eq:0416072313}
P\left[ k^{1/2} \Delta^\dagger \ge c_0 u \right]
 \le { \frac{p K_\theta^\theta}  {(\kappa u)^{\theta/2} } },
\end{eqnarray}
where $c_0$ is a constant only depending on $\theta$, and the distributional distance
\begin{equation}\label{eq:0416032335}
\rho^\dagger \lesssim 
 { k^{-1/6} (\log p)^{7/6} 
 + (p k^{-\theta/4})^{2/(\theta+2)} (\log p)^{(3\theta-4) /(4 + 2\theta)},}
\end{equation}
where the constant in $\lesssim$ only depends on $\theta$, $\kappa$ and $K_\theta$. 
\end{corol}

Corollary \ref{th:0416032326} follows from Theorem \ref{th:0416032233} by replacing $\theta$ in the latter by $\theta/2$ in view of (\ref{eq:0416032319}).  In comparison with (\ref{eq:0416082200}), Corollary \ref{th:0416032326} requires the stronger condition 
{ $p^{4/\theta} (\log p)^{3-4/\theta} = o(k)$} to ensure that $\rho^\dagger = o(1)$.

Estimation of $\sigma^2$ and $\kappa$ is discussed in Section \ref{sec:varest}. Recall that $g_{1-\alpha}$ is the $(1-\alpha)$th quantile of $\max_{0\le j \le p-k}G_j^{\circ}$, $\alpha \in (0,1)$. Corollary \ref{th:0416032326} suggests rejecting $H_0$ and accepting the alternative hypothesis $H_1$ of (\ref{eq:0416031319}) at level $\alpha$ if 
$
\max_{0\le j \le p-k}R_j^{\dagger} > \hat{\kappa} g_{1-\alpha}.
$

\begin{remark}  
\label{rem:1}
Denote by $\Delta^\circ_k$ the quantity $\Delta^\circ$ in (\ref{eq:0416071927}). A careful analysis of the proof of Theorem \ref{th:KC} (which implies Theorem \ref{th:0416032233} with $n = 1$) indicates that Theorem \ref{th:0416032233} is still valid with $\Delta^\circ_k$ (resp. $R_k^\bullet := \sqrt k \max_{0 \le j \le p-k} R_{j}^\circ / \sigma$ and $G_k^\bullet := \sqrt k \max_{0 \le j \le p-k} G^\circ_j$) therein replaced by the uniform version $\max_{k \le m \le p} \Delta^\circ_m$ (resp. $\max_{k \le m \le p} R_m^\bullet$ and $\max_{k \le m \le p} G_m^\bullet$). Similarly, for the two-sided test, Corollary \ref{th:0416032326} also holds with the uniform version $\max_{k \le m \le p} m^{-1/2} \max_{0 \le j \le p-m}$\newline $\sum_{i=j+1}^{j+m} (X_i^2 - \sigma^2) $. The latter quantity has an interesting connection with the adaptive Neyman's high dimensional multivariate normal mean test which has the form $\max_{1 \le m \le p} (2m)^{-1/2} \sum_{i=1}^{m}$ \newline $(X_i^2 - \sigma^2) $, which was considered in Section 2.1 in \cite{fan96} in the setting that large values of $\mu$ concentrate on the first $m$ dimensions and $X_i \overset{i.i.d.}{\sim} N(\mu_i, 1)$. Here $m$ is estimated by the maximizer  $\hat m = {\rm argmax}_{1\le m \le p} (2m)^{-1/2} \sum_{i=1}^{m} (X_i^2 - \sigma^2) $.
\end{remark}

\subsection{An Algorithm for Locating Break-points}
\label{algorithm}
Once the null hypothesis is rejected, we need to locate break-points. We propose Algorithms \ref{algorithm:onesided} and \ref{algorithm:twosided} for locating break-points based on the one- and the two-sided tests, respectively. 
Theoretical properties of Algorithm \ref{algorithm:onesided} (resp. \ref{algorithm:twosided}) are given in Theorem \ref{th:locations} (resp. \ref{th:two-locations}).

\subsubsection{Locating break-points based on one-sided test}
We first present an algorithm based on the one-sided test.
\begin{algorithm}
\label{algorithm:onesided} 
Step 1. Let $L_j^{\circ} = R_{j-k}^{\circ}, j = k, \ldots, p$. Compute $Q_j^{\circ} = 1(R_j^{\circ} > \gamma) + 1(L_j^{\circ} > \gamma)$ for a pre-specified cutoff value $\gamma$, $j = k, \ldots, p-k.$ We use a majority vote approach to smooth $Q_j^{\circ}.$ Specifically, denote $j_0 = \sum_{j=i-k}^{i+k}I\{Q_j^{\circ} = 0\}, j_1 = \sum_{j=i-k}^{i+k}I\{Q_j^{\circ} = 1\},$ and $j_2 = \sum_{j=i-k}^{i+k}I\{Q_j^{\circ} = 2 \}.$ Let $\tilde{Q}_j^{\circ} = \{k, \mbox{such that} \ j_k = \max_{l\in \{0,1,2\}} j_l\}.$

Step 2. Decompose $\{1, \ldots, p\} = W_0 \cup W_1 \cup W_2,$ where $j \in W_0$ if $\tilde{Q}_j^{\circ} =0, j \in W_1$ if $\tilde{Q}_j^{\circ}=1$ and $j \in W_2$ if $\tilde{Q}_j^{\circ}=2.$ Let ${\cal M}_1, \ldots, {\cal M}_{\hat{l}}$ be connected components of $W_1.$

Step 3. Given $\delta < \gamma$, the break-points are defined as $\hat{\tau}_i = \mbox{argmax}_{j \in {\cal M}_i}\{R_j^{\circ}: \, L_j^{\circ} \le \delta\}$ if ${\cal M}_i$ is the transition region from $W_0$ to $W_2$. If ${\cal M}_i$ is the transition region from $W_2$ to $W_0,$ $\hat{\tau}_i = \mbox{argmax}_{j \in {\cal M}_i}\{L_j^{\circ}: \, R_j^{\circ} \le \delta\}$.
\end{algorithm}

{The estimated signal sets are $\hat {\cal S}_1 = \{ \hat \tau_1, \ldots, \hat \tau_2 - 1\}, \hat {\cal S}_3 =  \{ \hat \tau_3, \ldots, \hat \tau_4 - 1\}, \ldots$.} The rationale behind Algorithm \ref{algorithm:onesided} is that if $\mu_j = 0$, then $R_j^{\circ}$ is close to $0;$ on the other hand, in the signal clusters, $R_j^{\circ}$ tends to be large. By locally averaging the data, we can reduce the variability, which has the effect of boosting the signal noise ratio. If there are many weak signals, we are able to detect them by the aggregation. On the other hand, if sporadic large values of $X_j$ arise, they can be smoothed out through $R_j^{\circ}$ to avoid false discoveries. Therefore, we can effectively de-noise the data to achieve better inference. In the signal cluster, $Q_j^{\circ}$ is most likely to be $2$ and in the non-signal cluster, $Q_j^{\circ}$ is most likely to be $0.$ In the boundary between signal and non-signal cluster, $Q_j^{\circ}$ is most likely to be $1.$ After Step 1, we get smoothed $\tilde{Q}_j^{\circ}$ that are in clusters of $0, 1$ and $2.$ Step 2 focuses on the signal and non-signal cluster boundary regions, where $\tilde{Q}_j^{\circ}=1.$ Step 3 locates break-points. The basic idea is that without noise at the true break-points, $R_j^{\circ}$ reaches the maximum as there is no noise to dilute the summation if we are transiting from non-signal cluster to signal cluster. The constraint $L_j^{\circ}\le \delta$ prevents the detected break-points to be too far from the true break-points when $\mu_j$ increases in the signal cluster. Similarly, without noise, $L_j^{\circ}$ obtains the maximum if we are transiting from signal to non-signal cluster at the true break-points. The constraint $R_j^{\circ} \le \delta$ prevents the detected break-points to be too far from the true break-points when $\mu_j$ decreases in the signal cluster. With two thresholds $\delta < \gamma$, Algorithm \ref{algorithm:onesided} has more flexibility and produces more accurate estimates of the break-points than the procedure in \cite{caowu15} which only uses one threshold $\gamma$.

Our method depends on the choice of window size $k$ and thresholds $\gamma$ and $\delta$. Theoretically speaking, the allowable range of $k$ is specified in (\ref{eq:0416082200}). Our simulation studies show that the proposed method is relatively robust to different choices of $k.$ In practice, following the idea of the adaptive Neyman's high dimensional multivariate normal mean test mentioned in Remark \ref{rem:1}, as a simple rule of thumb choice we can let $\hat m = \argmax_{m \ge \sqrt p} R_m^\bullet$ and $k = \lfloor \hat m / 2 \rfloor$. For a data-driven selection of $\gamma$ and $\delta$, we can choose $\gamma = \hat \sigma g_{1-\alpha}$ and $\delta = \hat \sigma g_{1, 1-\alpha}$, where $g_{1-\alpha}$ and $g_{1,1-\alpha}$ are the $(1-\alpha)$th quantiles of $\mbox{max}_{0\le j\le p-k} G_j^{\circ}$ and $\mbox{max}_{j\in W_1 } G_j^{\circ}$, respectively, with $\alpha$ close to $0$. They can be obtained by simulations. Section \ref{sec:varest} gives an estimate $\hat \sigma$ of $\sigma$.

\begin{condition}
\label{cond:one-sided-H1}
Recall ${\cal S}_f = \{ \tau_f, \ldots, \tau_{f+1}-1\}$ and ${\cal S} = {\cal S}_1 \cup {\cal S}_3 \cup \ldots$ is the signal set. Assume $d := \min_{i \in {\cal S}} \mu_i > 0$ and $2 k < \min_f (\tau_{1+f} - \tau_f)$.
\end{condition}

\begin{condition}
\label{cond:subG}
We say that a random variable $Z$ is $\sigma^2$-{\it sub-Gaussian} if $E (\exp(u Z/\sigma ) ) \le \exp(u^2/2)$ for all $u \in {\mathbb R}$. Note that $N(0, \sigma^2)$ is $\sigma^2$-sub-Gaussian.
\end{condition}

To state Theorem \ref{th:locations}, we need to introduce truncated moment functions. For a random variable $X$ with $E (X^2) < \infty$, define the truncated moment 
\begin{equation}
\label{eq:08041839}
{\cal M}_\upsilon(X) = E \min(|X|^\upsilon, \, X^2)  < \infty, \,\, \upsilon > 2.
\end{equation}
If $X$ has finite $\theta$th moment with $2 < \theta < \upsilon$, then ${\cal M}_\upsilon(X) \le E(| X |^\theta)$. Theorem \ref{th:locations}(i) (resp. (ii)) concerns sub-Gaussian (resp. polynomial-tailed) noises.

\begin{thm}
\label{th:locations}
Assume Condition \ref{cond:one-sided-H1} and $d/2 > \gamma > \delta$. (i) Assume Condition \ref{cond:subG} holds for $Z_j$. {Denote by $\hat{l}$ the estimated number of break points.} Then
\begin{eqnarray}\label{eq:subGclose}
1-P\left[ \hat l = l, \, \max_{j \le l} |\hat \tau_j - \tau_j| \le {2 k \delta \over d} \right]
 \le c_3( {p \over k} e^{-c_1 k \gamma^2 / \sigma^2} + l e^{-c_2 k \delta^2 / \sigma^2}),
\end{eqnarray}
where $c_1, c_2, c_3$ are absolute constants. (ii) Assume Conditions \ref{cond:0416032220} and let $\upsilon \ge \theta$. Then
\begin{eqnarray}\label{eq:06021622}
1-P\left[ \hat l = l, \, \max_{j \le l} |\hat \tau_j - \tau_j| \le {2 k \delta \over d} \right]
  &\lesssim&  p {\cal M}_\upsilon( Z_1 / (k \gamma)) + {p \over k} e^{-c_1 k \gamma^2 / \sigma^2} \cr
  & &+ l k {\cal M}_\upsilon( Z_1 / (k \delta)) + l e^{-c_2 k \delta^2 / \sigma^2} \cr
  &\le&  
  K_\theta^\theta (p \gamma^{-\theta} + l k \delta^{-\theta})  k^{-\theta}  \cr
  & &+ p k^{-1} e^{-c_1 k \gamma^2 / \sigma^2} + l e^{-c_2 k \delta^2 / \sigma^2},
\end{eqnarray}
where $c_1$ and $c_2$ are absolute constants and the constant in $\lesssim$ only depends on $\theta$ and $\upsilon$.
\end{thm}

Theorem \ref{th:locations} is proved in the Supplementary Material. In comparison with (\ref{eq:subGclose}), the extra term $K_\theta^\theta (p \gamma^{-\theta} + l k \delta^{-\theta})  k^{-\theta}$ in (\ref{eq:06021622}) is due to polynomial tails, which are heavier than the sub-Gaussian ones. In the sub-Gaussian case (i) with unbounded $l$ (namely $l \to \infty$), choose $\gamma = C_1 (k^{-1} \log p)^{1/2}$, and $\delta = C_2 (k^{-1} \log l)^{1/2}$, where $C_1$ and $C_2$ are sufficiently large constants, we have the uniform bound $\max_{j \le l} |\hat \tau_j - \tau_j| = {O_P(d^{-1} (k \log l)^{1/2})}$. The condition $d/2 > \gamma$ requires that $k \ge C_3 d^{-2} \log p$ for a sufficiently large constant $C_3$. When $l$ is bounded, { by letting $k=\lfloor Cd^{-2}\log p \rfloor$ for a sufficiently large $C$,}  we can similarly obtain the uniform bound $\max_{j \le l} |\hat \tau_j - \tau_j| = {O_P(d^{-2} (\log p)^{1/2})}$. In the context of detecting a deterministic signal with unknown spatial extent in the univariate sampled data model with standard white Gaussian noises, \cite{chan2013detection} dealt with the special case $\mu_j = d {\bf 1}_{\tau_1 \le j < \tau_2}$ and considered the consistency of detection based on scan statistics under the condition $\tau_2 - \tau_1 \ge c_p d^{-2} \log p$, where $c_p = 2 + \iota_p$ and $\iota_p^2 \log p \to \infty$. The latter observation has a similar flavor as our condition $k \ge C_3 d^{-2} \log p.$ 

The polynomial-tailed case (\ref{eq:06021622}) is more involved. To ensure that the right hand side of (\ref{eq:06021622}) goes to $0$, we can choose $\gamma = C_1 (k^{-1} p^{1/\theta} + (k^{-1} \log p)^{1/2})$ and $\delta = C_2 (k^{-1} (l k)^{1/\theta} + (k^{-1} \log l)^{1/2})$, where $C_1, C_2 > 0$ are sufficiently large constants. Assume $k \ge C_3 ( d^{-2} \log p + d^{-1} p^{1/\theta})$ for a sufficiently large constant $C_3$, 
we have the uniform consistency $\max_{j \le l} |\hat \tau_j - \tau_j| \le O_P( k \delta/ d)$. Thus the numbers of false discoveries and missed discoveries are bounded by $l O_P(k \delta / d)$. If $l$ is bounded, then the latter bound becomes $O_P[(k \log p)^{1/2}) / d]$.



\subsubsection{Locating break-points based on two-sided test}

We next present an algorithm based on the two-sided test. It is similar to Algorithm \ref{algorithm:onesided}. With the square form (\ref{eq:0316302218}), it can pick up signals with alternating positive and negative signs. Same simulation assisted choice of $\gamma$ and $\delta$ as in the one-sided test can be used.

\begin{algorithm}
\label{algorithm:twosided}
Step 1: Calculate $R_i^{\dagger}$ and let $L_i^{\dagger} = R_{i-k}^{\dagger},i = k, \ldots, p.$ For a pre-specified $\gamma,$ let $Q_i^{\dagger} = 1(R_i^{\dagger} > \gamma) + 1(L_i^{\dagger} > \gamma), i =k, \ldots, p-k.$ The same majority vote approach as in Algorithm \ref{algorithm:onesided} is used to smooth $Q_i^{\dagger},$ denoted as $\tilde{Q}_i^{\dagger}.$ 

Step 2: Decompose $\{1, \ldots, p\} = W_0 \cup W_1 \cup W_2,$ where $i\in W_0$ if $\tilde{Q}_i^{\dagger} =0, i\in W_1$ if $\tilde{Q}_i^{\dagger}=1$ and $i\in W_2$ if $\tilde{Q}_i^{\dagger}=2.$ Let ${\cal M}_1, \ldots, {\cal M}_{\hat{l}}$ be connected components of $W_1.$ 

Step 3. Given $\delta < \gamma$, the break-points are estimated as $\hat{\tau}_i = \mbox{argmax}_{j \in {\cal M}_i}\{R_j^{\dagger}: \, L_j^{\dagger} \le \delta\}$ if ${\cal M}_i$ is the transition region from $W_0$ to $W_2$. If ${\cal M}_i$ is the transition region from $W_2$ to $W_0,$ $\hat{\tau}_i = \mbox{argmax}_{j \in {\cal M}_i}\{L_j^{\dagger}: \, R_j^{\dagger} \le \delta\}$.
\end{algorithm}

 
\begin{condition}
\label{cond:two-sided-H1}
Recall Condition \ref{cond:one-sided-H1} for ${\cal S}$. Let $d = \min_{i \in {\cal S}} |\mu_i| > 0$ and assume $2 k < \min_f (\tau_{1+f} - \tau_f)$.
\end{condition}

\begin{thm}
\label{th:two-locations}
Assume Conditions \ref{cond:subG}, \ref{cond:two-sided-H1}, and $(k^{-1} \log p)^{1/2} = o(d^2)$. Let $\gamma = c_1 (k^{-1} \log p)^{1/2}$ and $\delta = c_2 (k^{-1} \log l)^{1/2}$, where $c_1$ and $c_2$ are sufficiently large constants. Then there exists a constant $c > 0$ independent of $k$ and $p$ such that  
\begin{eqnarray}
\label{eq:06031046}
P\left[ \hat l = l, \max_{j \le l} |\hat \tau_j - \tau_j| \le {c k \delta \over d^2} \right] \to 1.
\end{eqnarray}
\end{thm}

Theorem \ref{th:two-locations} provides a bound for uniform deviations of the estimated break-points. It is proved in the Supplementary Material, where the polynomial-tailed case is also studied. Same choice of $\gamma$ and $\delta$ can be used as in the one-sided test scenario. 


\subsection{Variance Estimation}

\subsubsection{Estimation of $\sigma^2$}
\label{sec:varest}
To apply Theorem \ref{th:0416032233} and Corollary \ref{th:0416032326} for computing the cutoff values based on $R_j^\circ$ and $R_j^\dagger$, we need to deal with the key issue of estimating the variance $\sigma^2$. Furthermore, to use $R_j^\dagger$, we need to estimate $\kappa^2$. For the nonparametric regression model
$X_i = \mu_i + Z_i = f(i/p) + Z_i, \, 1 \le i \le p$,
where $\mu_i = f(i/p)$, $f$ is a smooth function and $Z_i$ are i.i.d. with mean $0$ and variance $\sigma^2$, the problem of estimating $\sigma^2$ has a long history; see \cite{hkt90} and references therein. However the difference-based method in the latter paper does not work here. Due to the presence of the nonzero $\mu_j$s, the problem of estimating $\sigma^2$ is highly nontrivial. The latter problem is further complicated by the fact that the nonzero $\mu_j$s in the signal segments can change wildly. Here we shall use order statistics and obtain a consistent estimator. Let 
\begin{equation}
\label{eq:04112214}
\hat{\sigma}^2_i = \frac{1}{m}\sum_{j=i}^{i+m-1} X_j^2, \quad 1\le i \le p', \mbox{ where } p' = p-m+1.
\end{equation}
Let $\hat{\sigma}^2_{(1)} \le \hat{\sigma}^2_{(2)} \le \ldots \le \hat{\sigma}^2_{(p')}$ be the order statistics of $\hat{\sigma}^2_{1}, \ldots, \hat{\sigma}^2_{p'}$. Theorem \ref{sigma-est} shows that, for any $k \le p'/2$, $\hat{\sigma}^2_{(k)}$ is a consistent estimator of $\sigma^2$ under suitable conditions of $m$. The intuition is as follows: for large $m$, we expect that $\hat{\sigma}^2_{i} \approx E \hat{\sigma}^2_{i}  = \sigma^2 + m^{-1} \sum_{j=i}^{i+m-1} \mu_j^2$. The latter uniform closeness relation will be made rigorous in the proof of Theorem \ref{sigma-est}, which is proved in the Supplementary Material. Under Condition \ref{cond:null} below, we expect that majority of $\sum_{j=i}^{i+m-1} \mu_j^2$ will be $0.$ Thus the median or any lower quantile of $E \hat{\sigma}^2_{i}$ is $\sigma^2$.

In practice we choose the sample median estimate with $k = p'/2$.   

\begin{condition}
\label{cond:null}
There exists a constant $c > 0$ such that the length of non-signal clusters $\tau_{i+1} - \tau_{i} \ge c p$ for all even $i$, and the total length ${\sum_{i \ \mbox{is even}}}\ (\tau_{1+i} - \tau_{i}) \ge \lambda p$ with constant $\lambda > 1/2$.
\end{condition}

Condition \ref{cond:null} implies the natural requirement that the proportion of non-signals (namely $j$ with {$\mu_j \ne 0$}) is larger than $1/2$.

\begin{thm}\label{sigma-est}
Assume $(\mu_j)$ satisfies (\ref{eq:alt}), Condition \ref{cond:null}, $Z_i \in {\cal L}^\theta, \theta > 4$, $p^{2/\theta} = o(m)$ and $m = o(p)$. Then we have for any $k \le p'/2$ that
\begin{eqnarray}
\label{eq:15102235}
\hat{\sigma}^2_{(k)} = \sigma^2 + O_P(\gamma_p), 
 \mbox{ where } \gamma_p = ({ \log p \over m})^{1/2} + { p^{2/\theta} \over m}.
\end{eqnarray}
If Condition \ref{cond:subG} holds, $\log p = o(m)$ and $m = o(p)$, then $\hat{\sigma}^2_{(k)} = \sigma^2 + O_P(({m^{-1} \log p})^{1/2})$.
\end{thm}

\subsubsection{Estimation of $\kappa$}

The estimation of $\kappa$ in (\ref{eq:kappa}) is much more involved. The key issue is to estimate the fourth order moment $E (Z_i^4)$. Unlike (\ref{eq:04112214}), we cannot simply use order statistics of the moving window sample averages $\Xi_i := m^{-1} \sum_{j=i}^{i+m-1} X_j^4, 1\le i \le p-m+1,$ to estimate $E (Z_i^4)$, since the median or lower quantile of $E(\Xi_i) = E (Z_i^4) + m^{-1} \sum_{j=i}^{i+m-1} (\mu_j^4 + 6 \mu_j^2 \sigma^2 + 4 \mu_j^3 E (Z_j^3))$ is generally {\it not} $E (Z_i^4)$ if $E (Z_j^3) \not= 0$. $E(\Xi_i)$ can be greater or less than $E(Z_i^4)$ depending on what $\mu_j, j =i, \ldots, i+m-1$ and $E (Z_j^3)$ are. The reason is that the function $E(\mu + Z_j)^4$ may not be minimized at $\mu = 0$. For example, if $Z_j = E_j - 1$ with $E_j \sim \exp(1)$, then $E(\mu + Z_j)^4$ is minimized at $\mu \approx  -0.596072$. To circumvent the latter problem, we introduce
\begin{eqnarray}
\hat \nu_i = {1 \over m} \sum_{j=i}^{i+m-1} (X_j - X_{j-1})^4, \quad 2 \le i \le p-m+1,
\end{eqnarray}
and $\nu = E(Z_1 - Z_0)^4 = 2 \kappa^2 + 8 \sigma^4$. Note that $E(\mu + Z_1-Z_0)^4$ is indeed minimized at $\mu = 0$. The above estimate resembles the first order difference based estimate; see \cite{hkt90}. However the setting and the motivation are quite different. Let $\hat \nu_{(2)} \le \ldots \le \hat \nu_{(p-m+1)}$ be the order statistics. Corollary \ref{kappa-est} below concerns asymptotics for $\hat \nu_{(k)}$. It is proved in the Supplementary Material. Then we can estimate $\kappa^2$ by
 $\hat \kappa^2 = \hat \nu_{(k)} / 2 - 4 \hat{\sigma}^4_{(k)}$.
In practice we can choose $k = p' / 2$, which corresponds to the sample median. By Theorems \ref{sigma-est} and Corollary \ref{kappa-est}, we have $\hat \kappa^2 = \kappa^2 + O_P({\phi_{p,m}})$,
 {where $\phi_{p,m}$ is a function of $p$ and $m,$ given in the following corollary. }
\begin{corol}
\label{kappa-est}
Assume (\ref{eq:alt}), Condition \ref{cond:null} and that $Z_i \in {\cal L}^q, q > 4$, $p^{4/q} = o(m)$ and $m = o(p)$. Then we have for any $k \le p'/2$ that
\begin{eqnarray}
\label{eq:05281525}
\hat{\nu}^2_{(k)} = \nu + O_P({\phi_{p,m}}), 
 \mbox{ where } {\phi_{p,m}} = ({ \log p \over m})^{1/2} + { p^{4/q} \over m}.
\end{eqnarray}
If Condition \ref{cond:subG} holds, $(\log p)^2 = o(m)$ and $m = o(p)$, then $\hat{\nu}^2_{(k)} = \nu + O_P(({m^{-1} \log p})^{1/2})$.
\end{corol}


\vspace{-1.2cm}

\section{Test and estimation with multiple realizations}
\label{sec:multisep}
In Section \ref{sec:onereal}, only one realization $(X_j)_{j=1}^p$ is available, under the assumption that the errors $Z_j$ are i.i.d. When we have more than one realization, we will be able to detect clustered signals even if the variances change along the sequence. The allowance of heteroscedasticity substantially expands the application of our methods. Let $n ({\ge 2})$-realizations $Y_i = (Y_{i 1}, \ldots, Y_{i p})^T$ be observed, $i = 1, \ldots, n$, with 
\begin{eqnarray}
Y_{i j} = \mu_j + Z_{i j}, \,\, 1 \le j \le p,
\end{eqnarray}
where $Z_{i j}$ has mean $0$, variance $\sigma_j^2$ and independent across {both $i$ and $j$}. We are interested in testing (\ref{eq:0416031319}) and (\ref{eq:alt}). To this end, we propose a new test statistic and derive an omnibus test under the global null hypothesis $H_0$ in (\ref{eq:0416031306}).  Let $\hat \mu_j = n^{-1} \sum_{i=1}^n Y_{i j}$.

\subsection{One-sided Test}
Given a window size $k$, define
\begin{equation}\label{eq:0316241114}
R_j^\star =  { {\sum_{l=j+1}^{j+k} \sqrt n \hat \mu_l} \over {v_j^{1/2}} }, \, \mbox{ where }
 v_j = \sum_{l=j+1}^{j+k} \sigma^2_l,
\,   0 \le j \le p-k.
\end{equation}
Let $(G^\star_j)_{0 \le j \le p-k}$ be a mean zero Gaussian vector which has the same covariance structure as $(R^\star_j)_{0 \le j \le p-k}$. As a stochastic realization, we can let
\begin{equation}\label{eq:0316211620}
G^\star_j =  { W_j \over {v_j^{1/2}} }, \mbox{ where }
 W_j = \sum_{l=j+1}^{j+k} \sigma_l \eta_l, \, {v_j = E(W_j^2)} \ \mbox{ and } \eta_l \overset{i.i.d.}{ \sim} N(0, 1).
\end{equation}
Let $\sigma = (\sigma_1, \ldots, \sigma_p)$. Then $G^\star_j$ has marginal variance $1$ and covariance matrix $\Gamma(\sigma) = (\gamma_{j, j'}(\sigma))_{0 \le j, j' \le p-k}$ with $\gamma_{j, j'}(\sigma) = v^{-1/2}_j v^{-1/2}_{j'} E(W_j W_{j'}). $
Note that $\gamma_{j, j'}(\sigma) = 0$ if $|j - j'| \ge k$ and $(W_j)$ are $(k-1)$-dependent. Let the coupled distance
\begin{eqnarray}\label{eq:0416072316}
\Delta^\star = \max_{0 \le j \le p-k} |R^\star_j - G^\star_j|.
\end{eqnarray}

Theorem \ref{th:KC} below concerns the Gaussian approximation in terms of the closeness of $R_j^\star$ and $G_j^\star$ with various metrics. It is proved in the Supplementary Material. Relation (\ref{eq:0416040913}) is a coupling statement which provides a tail probability inequality for the maximum distance $\Delta^\star$ on some common probability space, while (\ref{eq:rho}) is for the distributional distance
\begin{eqnarray}
\rho^\star :=
\sup_u |P(\max_{0 \le j \le p-k} R^\star_j \ge u) - P(\max_{0 \le j \le p-k} G^\star_j \ge u)|.
\end{eqnarray}
We shall impose the following regularity condition.

\begin{condition}
\label{cond:1}
Let $\theta > 2$. Assume that there exist {positive} constants $\sigma_*$, $\sigma^*$ and $K_\theta$ such that, for all $1 \le j \le p$, $\sigma_* \le \sigma_j \le \sigma^*$, and $E|Z_{i j}|^\theta \le K_\theta^\theta$.
\end{condition}


{ {
\begin{thm}
\label{th:KC}
Assume Condition \ref{cond:1} and $\mu_i = 0, 1\le i \le p$. (i) Let $\theta > 2$. Then there exists a Gaussian process $(G^\star_j)_{0 \le j \le p-k}$ such that on a possibly larger probability space, for all $u > 0$,
\begin{eqnarray}
\label{eq:0416040913}
P\left[ (n k)^{1/2} \Delta^\star \ge c_0 u \right]
 \le { {n p K_\theta^\theta} \over u^\theta \sigma_*^\theta },
\end{eqnarray}
where $c_0$ is a constant only depending on $\theta$. (ii) Let $\theta > 3$. Then the distributional distance
\begin{equation}\label{eq:rho}
{\rho^*} \lesssim  (n k)^{-1/6} (\log p)^{7/6} + (np / (nk)^{\theta/2})^{1/(\theta+1)} (\log p)^{(3 \theta-2) /(2+2\theta)},
\end{equation}
where the constant in $\lesssim$ only depends on $\theta$, $\sigma_*$, $\sigma^*$ and $K_\theta$. 
\end{thm}
}
}

We emphasize that our theorem does not require $n \to \infty$ and it is also applicable when $n$ is finite. For example, when $n = 2$ observations are available, if we choose the window size $k$ be sufficiently large such that {$p (\log p)^{ 3 \theta/2-1} = o(k^{\theta/2})$}, then by (\ref{eq:rho}) and elementary manipulations we still have ${\rho^*} \to 0$. Under the slightly weaker condition $p^{2/\theta} = o(k)$, $R_j^{\star}$ and $G_j^{\star}$ are uniformly close to each other in the sense of $\max_{k \le j \le p-k} |R_j^{\star} - G_j^{\star}| = o_P(1)$ in view of (\ref{eq:0416040913}).

\subsection{Calculating cutoff values}
\label{sec:cutoff}
If the variances $\sigma_j^2$ are known, given the level $0 < \alpha < 1$, we can choose the cutoff value $u = u_{1-\alpha}$ such that 
\begin{eqnarray}\label{eq:03161609}
 P(\max_{0 \le j \le p-k} G_j^{\star} \ge u_{1-\alpha}) = \alpha.
\end{eqnarray}
The above can be done by Monte Carlo simulations. 
Assuming that $p, n, k$ satisfy the relation
$
{(n p)^{2/\theta} (\log p)^{3-2/\theta} = o(n k).}
$
Then the right hand side of (\ref{eq:rho}) goes to $0$. By Theorem \ref{th:KC}, the test {$\max_{0 \le j \le p-k} R_j > u_{1-\alpha}$} has the asymptotically correct size $\alpha$.

\subsubsection{Estimation of block sum variances}

In general, however, the variances $\sigma_j^2$ are not known. Since we have multiple realizations, we can estimate them by the classical unbiased variance estimate
\begin{eqnarray}\label{eq:03160948}
\hat \sigma_j^2 = {1\over {n-1}} \sum_{i=1}^n (Y_{i j} - \hat \mu_j)^2.
\end{eqnarray}
Correspondingly, our test statistic $R_j^\star$ in (\ref{eq:0316241114}) now becomes
\begin{eqnarray}\label{eq:03160951}
\hat R_j =  { {\sum_{l=j+1}^{j+k} \sqrt n \hat \mu_l} \over {(\sum_{l=j+1}^{j+k} \hat \sigma^2_l)^{1/2}} }, \,\,   0 \le j \le p-k.
\end{eqnarray}
At first glance, if $n$ is small, $\hat \sigma_j^2$ may deviate substantially from $\sigma_j^2$. For example, if $n = 2$, then $\hat \sigma_j^2 = (Y_{1 j} - Y_{2 j})^2 / 2,$ which may be quite different from $\sigma_j^2$. This difference might suggest that replacing $\sigma_l^2$ in $R_j$ by $\hat \sigma_l^2$ can be problematic. However, interestingly, under suitable conditions on $p, k, n$, $R_j^{\star}$ and $\hat R_j$ can still be uniformly close. This can be intuitively explained by the fact that, in $R_j$, it is the block sum variance 
$
v_j = \sum_{l=j+1}^{j+k} \sigma^2_l
$
that is directly involved, not just a single $\sigma_l^2$. The sum 
$
\hat v_j = \sum_{l=j+1}^{j+k} \hat \sigma^2_l
$
can still be a good estimate of \textcolor{red}{$v_j$}, despite that individually the difference $\hat \sigma_j^2 - \sigma_j^2$ can be big due to a small $n$. The convergence rate is given in the following Proposition \ref{prop:03161008}. It implies that, under Condition \ref{cond:1}, if $p = o(n^{\theta/2-1} k^{\theta/2})$, then $\hat v_j / v_j$ is uniformly close to $1$. It is proved in the Supplementary Material. 

\begin{proposition}
\label{prop:03161008}
Let Condition \ref{cond:1} be satisfied. If $\theta > 4$, we have
\begin{equation}\label{eq:03161009}
P(n \max_{0 \le j \le p-k} |\hat v_j - v_j| > u )
 \lesssim  {{  n p K_\theta^\theta } \over { u^{\theta/2}} } 
  + {p \over k} \exp(- c_3 { u^2 \over {n k K_4^4}} ),
\end{equation}
where the constant in $\lesssim$ and $c_3 > 0$ only depend on $\theta$. If $2 < \theta \le 4$, then
\begin{eqnarray}\label{eq:03161103}
P(n \max_{0 \le j \le p-k} |\hat v_j - v_j| > u )
 \lesssim {{  n p K_\theta^\theta } \over { u^{\theta/2}} }. 
\end{eqnarray}
\end{proposition}

Note that (\ref{eq:03161009}) of Proposition \ref{prop:03161008} implies that we have the uniform convergence rate
\begin{equation*}
    n \max_{0 \le j \le p-k} |\hat v_j - v_j| = O_P( (n p)^{2/\theta} + (n k)^{1/2} \log p). 
\end{equation*}
Under Condition \ref{cond:1}, $k \sigma_*^2 \le v_j \le k (\sigma^*)^2$. Thus the term $n \max_{0 \le j \le p-k} |\hat v_j - v_j| $ in (\ref{eq:03161009}) can be replaced by the ratio normalized version $n k \max_{0 \le j \le p-k} |\hat v_j / v_j - 1|$ so that (\ref{eq:03161009}) is still valid with the constants in $\lesssim$ and $c_3$ therein depending on $\theta$, $\sigma_*$ and $\sigma^*$. By elementary calculations, if $p = o(n^{\theta/2-1} k^{\theta/2})$, the ratios $\hat v_j / v_j$ are uniformly close to $1$ in the sense that $\max_{0 \le j \le p-k} |\hat v_j / v_j - 1| = o_P(1)$.

\subsubsection{A bootstrap calibration procedure}

To perform the test for $H_0: \mu_j \equiv 0$ based on $\hat R_j$ with $\sigma_j^2$ replaced by their estimates $\hat \sigma_j^2$, we need to estimate the corresponding cutoff value $u_{1-\alpha}$ based on (\ref{eq:03161609}). Recall that $\Gamma(\sigma) = (\gamma_{j, j'}(\sigma))_{k \le j, j' \le p-k}$ is the covariance matrix for the vector $(Z_j)_{k \le j \le p-k}$. Write $u_{1-\alpha} = q_\alpha(\sigma)$ as a function of $\sigma = (\sigma_1, \ldots, \sigma_p)$. Write $\hat \sigma = (\hat \sigma_1, \ldots, \hat \sigma_p)$ and $\hat u_{1-\alpha} = q_\alpha(\hat \sigma)$ which satisfies
\begin{eqnarray}
\label{eq:03161616}
 P^*(\max_{0 \le j \le p-k} G^*_j \ge \hat u_{1-\alpha}) = \alpha,
\end{eqnarray}
where $P^*$ is the probability measure given $Y = (Y_1, \ldots, Y_n)$ and, given $\hat \sigma$, $(G^*_j)_{k \le j \le p-k}$ is mean $0$ Gaussian vector with covariance matrix $\Gamma(\hat \sigma)$. In particular, as (\ref{eq:0316211620}), we can define 
\begin{eqnarray}\label{eq:0316211621}
G^*_j =  { W_j^* \over {\hat v_j^{1/2}} }, 
\mbox{ where } W_j^* = {\sum_{l=j+1}^{j+k} \hat \sigma_l \eta_l}
\end{eqnarray}
and $\eta_l, l \in \mathbb Z$, are i.i.d. $N(0, 1)$ random variables that are independent of $Y = (Y_1, \ldots, Y_n)$. Given $\hat \sigma$, the cutoff value $\hat u_{1-\alpha}$ in (\ref{eq:03161616}) can also be computed by extensive simulations.

The following theorem shows the validity of the above plug-in method in the sense that the size of our test is close to $\alpha$. It is proved in the Supplementary Material. 

\begin{thm}
\label{th:0316212031}
Let  $t_1 = (\log p)^{1/2} (n k)^{-1/2}$,
$
t_2 = ( pn (n k)^{-\theta/2} (\log p)^{-2/3} )^{1/(1/3+\theta/2)}
$ and $t_* = \max(t_1, t_2)$.
Recall (\ref{eq:rho}) for $\rho^*$. Let $0 < \alpha < 1$. Then under Condition \ref{cond:1}, we have
\begin{eqnarray}\label{eq:0316212030}
\quad | P(\max_{0 \le j \le p-k} \hat R_j \ge \hat u_{1-\alpha}) - \alpha |
 \lesssim \rho^* + t_*^{1/3} (\log p)^{2/3},
\end{eqnarray}
where the constant in $\lesssim$ only depends on $\sigma_*, \sigma^*, \theta$ and $K_\theta$. 
In particular, the right hand side of (\ref{eq:0316212030}) is $o(1)$ if
$
pn  (\log p)^{3 \theta/2-1}  = o((n k)^{\theta/2}).
$
\end{thm}

\subsection{Estimating break-points based on one-sided test}

Algorithm \ref{algorithm:multionesided} shows estimating break-points based on the one-sided test. It uses $R_j^{\star}$ assuming that $\sigma_j$, $1 \le j \le p$, are known. If not known, we shall use the estimates $\sigma_j^2$ in (\ref{eq:03160948}). Same simulation assisted $\gamma$ and $\delta$ can be used as in the one realization one-sided test case. 
Theorem \ref{th:multione-sided} provides theoretical properties of the break-point estimates.

\begin{algorithm}
\label{algorithm:multionesided}

Step 1. Let $L_i^{\star} = R_{i-k}^{\star}, i = k, \ldots, p-k$ and denote $Q_i^{\star} = 1(R_i^{\star} > \gamma) + 1(L_i^{\star} > \gamma)$ for a pre-specified cutoff value $\gamma$. We use a majority vote approach to smooth $Q_i^{\star}.$ Specifically, denote $j_0^{\star} = \sum_{i=j-k}^{j+k}I\{Q_i^{\star} =0\}, j_1^{\star} = \sum_{i=j-k}^{j+k}I\{Q_i^{\star} =1\},$ and $j_2^{\star} = \sum_{i=j-k}^{j+k}I\{Q_i^{\star}=2\}.$ Let $\tilde{Q}_j^{\star} = \{k, \mbox{such that}\ j_k^{\star} = \mbox{max}_{l \in \{0, 1, 2 \}}j^{\star}_l \}.$

Step 2. Decompose $\{1, \ldots, p\} = W_0 \cup W_1 \cup W_2,$ where $i\in W_0$ if $\tilde{Q}_i^{\star} =0, i\in W_1$ if $\tilde{Q}_i^{\star}=1$ and $i\in W_2$ if $\tilde{Q}_i^{\star}=2.$ Let ${\cal M}_1, \ldots, {\cal M}_{\hat{l}}$ be connected components of $W_1.$

Step 3. Let $R_j^\flat = { {\sum_{f = j+1}^{j+k} \sqrt n \hat \mu_f} / \sqrt k}$ and $L_j^\flat = R_{j-k}^\flat$. Given $\delta < \gamma$, the break-points are defined as $\hat{\tau}_i = \mbox{argmax}_{j \in {\cal M}_i}\{R_j^{\flat}: \, L_j^{\star} \le \delta\}$ if ${\cal M}_i$ is the transition region from $W_0$ to $W_2$. If ${\cal M}_i$ is the transition region from $W_2$ to $W_0,$ $\hat{\tau}_i = \mbox{argmax}_{j \in {\cal M}_i}\{L_j^{\flat}: \, R_j^{\star} \le \delta\}$.
\end{algorithm}

Differently from Algorithm \ref{algorithm:onesided}, in Step 3 of Algorithm \ref{algorithm:multionesided} we use $R_j^{\flat}$ instead of $R_j^{\star}$ in the argmax function.  The reason is for technical convenience: one has monotonicity $E(R_j^\flat) < E(R_i^\flat)$ for $\tau_1 - k < j < i \le \tau_1$, which tends to make the estimated break-point closer to $\tau_1$. In comparison $E(R_j^{\star})$ is generally not monotone, since the variances $\sigma^2_j$ can be unequal.

\begin{thm}
\label{th:multione-sided}
Assume Conditions \ref{cond:one-sided-H1}, \ref{cond:null}, $Z_{i j}$ are $\sigma^2$-sub-Gaussian, and $2 \sigma \gamma \le d \sqrt{n k}$. Let \newline $m = \lfloor 2 k^{1/2} \delta \sigma n^{-1/2} d^{-1} \rfloor$. Then
\begin{eqnarray}
\label{eq:06032216}
1 - P\left[ \hat l = l, \, \max_{j \le l} |\hat \tau_j - \tau_j| \le m \right]
 \lesssim {p \over k} \exp(-c_1 \gamma^2 ) + l \exp(-c_2 \delta^2 ) ,
\end{eqnarray}
where the constant in $\lesssim$ and $c_1, c_2 > 0$ are independent of $k, d, n$ and $p$.
\end{thm}


Theorem \ref{th:multione-sided} is proved in the Supplementary Material. Assume that $(\log p) (\log l) = o(n^2 d^4)$ and $k$ satisfies $(d^2 n)^{-1} \log p = o(k)$ and $k = o(n d^2 / \log l)$. Let $\gamma = C_1 (\log p)^{1/2}$, $\delta = C_2 (\log l)^{1/2}$, where $C_1, C_2 > 0$ are constants. Then the right hand side of (\ref{eq:06032216}) can be arbitrarily small by letting $C_1, C_2$ sufficiently large. Theorem \ref{th:multione-sided} implies that we can have exact recovery with probability $P[ \hat l = l, \, \max_{j \le l} |\hat \tau_j - \tau_j| = 0] \to 1$ since $k^{1/2} \delta n^{-1/2} d^{-1} \to 0$.

\vspace{-0.5cm}

\subsection{Two-sided test: A U-statistic approach}  
In the one-realization case, we use (\ref{eq:0316302218}) to test the two-sided alternative $H_1$ of (\ref{eq:alt}). If multiple realizations $Y_i = (Y_{i 1}, \ldots, Y_{i p})^T$, $1 \le i \le n$, are available, we can use the $U$-statistic
\begin{eqnarray}
\label{eq:Wj15}
W_j = {2 \over {n (n-1)}} \sum_{1 \le i < i' \le n} Y_{i j} Y_{i' j},
\end{eqnarray}
which is an unbiased estimate of $\mu_j^2$. This is different from (\ref{eq:0316302218}) in that $X_j^2$ in the latter is not an unbiased estimate of $\mu_j^2$. As an important consequence, we remark that unlike the two-sided test in Section \ref{sec:12}, here we do not need to use $\kappa$ of form (\ref{eq:kappa}). Under $H_0$, the variance of $W_j$ is $2 \sigma_j^4 / (n(n-1))$.  Define
\begin{eqnarray}
R_{j, 4} = ((n^2-n)/2)^{1/2} {{W_{j+1} + \ldots + W_{j+k}} 
 \over { (\sigma^4_{j+1} + \ldots + \sigma^4_{j+k})^{1/2} }}.
\end{eqnarray}
Let $\eta_i, i \in \mathbb Z,$ be i.i.d. $N(0, 1)$. Define the Gaussian process
\begin{eqnarray}
G_{j, 4} = {{\sigma_{j+1}^2 \eta_{j+1} + \ldots + \sigma_{j+k}^2 \eta_{j+k}} 
 \over { (\sigma^4_{j+1} + \ldots + \sigma^4_{j+k})^{1/2} }}.
\end{eqnarray}

\begin{thm}
\label{th:07231045}
Assume Condition \ref{cond:1} and $\mu_i = 0, 1\le i \le p$. Then the distributional distance 
\begin{eqnarray}
\label{eq:07231046}
& &
\sup_u |P( \max_{0 \le j \le p-k} R_{j, 4} \ge u) - P( \max_{0 \le j \le p-k} G_{j, 4} \ge u)| \cr
 & &\lesssim  k^{-1/6} (\log p)^{7/6} 
 + (p k^{-\theta/2})^{1/(\theta+1)} (\log p)^{(3 \theta-2)/(2+2\theta)}.
\end{eqnarray}
\end{thm}

In $R_{j, 4}$, the quantity $\sigma_j^4$ is typically unknown. Here we shall propose an unbiased estimate. Note that the natural estimate $(\hat \sigma_j^2)^2$ with $\hat \sigma_j^2$ given in (\ref{eq:03160948}) is not unbiased. Let
\begin{eqnarray}
\label{eq:0416062209}
\quad
\hat \omega_j = { {1/4} \over {n (n-1) (n-2) (n-3)}} \sum
(Y_{i j} - Y_{i' j})^2 (Y_{h j} - Y_{h' j})^2,
\end{eqnarray}
where the sum is over mutually different indexes $i, i', h, h' \in \{1, \ldots, n\}$. Clearly $E (\hat \omega_j) = \sigma_j^4$. Similar to (\ref{eq:03160951}), consider the realized version
\begin{eqnarray}
R^*_{j, 4} = ((n^2-n)/2)^{1/2} {{W_{j+1} + \ldots + W_{j+k}} 
 \over { (\hat \omega_{j+1} + \ldots + \hat \omega_{j+k})^{1/2} }}.
\end{eqnarray}
To test $H_0$ vs $H_1$ in (\ref{eq:alt}), we reject $H_0$ at level $\alpha \in (0, 1)$ if $\max_{0 \le j \le p-k} R^*_{j, 4} \ge q_{1-\alpha}$ for some cutoff value $q_{1-\alpha}$. As in (\ref{eq:03161616}), $q_{1-\alpha}$ can be approximated by $\hat q_{1-\alpha}$, which satisfies
$ P^*(\max_{0 \le j \le p-k} G^*_{j, 4} \ge \hat q_{1-\alpha}) = \alpha$, 
where  
\begin{eqnarray}
\label{eq:0416062054}
  G^*_{j, 4} = {{\hat \omega_{j+1}^{1/2} \eta_{j+1} + \ldots + \hat \omega_{j+k}^{1/2} \eta_{j+k}} 
 \over { (\hat \omega_{j+1} + \ldots + \hat \omega_{j+k})^{1/2} }}
\end{eqnarray}
a Gaussian process conditioning on $(Y_1, \ldots, Y_n)$.
As a slightly modified version, noting that for i.i.d. $N(0, 1)$ random variables $Z_1, \ldots Z_n$, the $U$-statistic $2 \sum_{1 \le i < i' \le n} Z_i Z_{i'} = (n^2-n) \bar Z_n^2 - \sum_{i=1}^n (Z_i - \bar Z_n)^2$ is identically distributed as $\zeta = (n-1) \chi^2_1 - \chi^2_{n-1}$, where the $\chi^2$ random variables $\chi^2_1$ and $\chi^2_{n-1}$ are independent, we can use 
\begin{eqnarray}
\label{eq:0416062057}
  G^\diamond_{j, 4} ={{\hat \omega_{j+1}^{1/2} \zeta_{j+1} + \ldots + \hat \omega_{j+k}^{1/2} \zeta_{j+k}} 
 \over { (\hat \omega_{j+1} + \ldots + \hat \omega_{j+k})^{1/2} }},
\end{eqnarray}
where $\zeta_i$ are {independent and identically distributed} as $\zeta/\sqrt{2n(n-1)}$. If $n$ is big, $G^\diamond_{j, 4}$ gives a better approximation. 

In the definition of $\hat \omega_j$ in (\ref{eq:0416062209}), it involves a 4-fold summation with $O(n^4)$ computation complexity. Interestingly, we can have the following expression which allows computing $\hat \omega_j$ within only $O(n)$ steps: elementary but tedious calculations show 
\begin{equation}\label{omegahat}
\hat \omega_j = {{ (n-1)(4 S_{j,3}S_{j,1} - n S_{j, 4} - 3 S_{j, 2}^2)
 + (n S_{j, 2} - S_{j,1}^2)^2}
\over {n (n-1) (n-2) (n-3)}}, \mbox{ where }  S_{j, l} = \sum_{i=1}^n Y_{i j}^l.
\end{equation}
To compute $W_j$ in (\ref{eq:Wj15}), we use the well-known formula
$
W_j = {{S_{j,1}^2 - S_{j,2}} /(n (n-1))}.
$

\subsubsection{Estimating break-points based on two-sided test}
Similar to Algorithm \ref{algorithm:twosided}, we can adjust Algorithm \ref{algorithm:multionesided} for locating break-points based on two-sided test. Theorem \ref{th:multitwo-sided} shows theoretical properties of Algorithm \ref{algorithm:multitwosided} and it is proved in the Supplementary Material. 

\begin{algorithm}
\label{algorithm:multitwosided}

Step 1. Let $Q_{j,4} = 1(R_{j,4} > \gamma) + 1(L_{j,4} > \gamma)$ for a pre-specified cutoff value $\gamma$, $j =k, \ldots, p-k.$ We use a majority vote approach to smooth $Q_{j,4}.$ Specifically, denote $l_0 = \sum_{i=j-k}^{j+k}I\{Q_{j,4}=0\}, l_1 = \sum_{i=j-k}^{j+k}I\{Q_{j,4}=1\},$ and $l_2 = \sum_{i=j-k}^{j+k}I\{Q_{j,4}= 2\}.$ Let $\tilde{Q}_{j,4} = \{k, \mbox{such that}\  l_k = \mbox{max}_{j \in \{0, 1, 2\}}l_j \}.$

Step 2. Decompose $\{1, \ldots, p\} = W_0 \cup W_1 \cup W_2,$ where $j \in W_0$ if $\tilde{Q}_{j,4} =0, j \in W_1$ if $\tilde{Q}_{j, 4} = 1$ and $j \in W_2$ if $\tilde{Q}_{j,4} = 2.$ Let ${\cal M}_1, \ldots, {\cal M}_{\hat{l}}$ be connected components of $W_1.$

Step 3. Let $R^\ddagger_j =  ((n^2-n)/2)^{1/2} \sum_{h=1}^k W_{j+h} / { k^{1/2} }$ and  
$L^\ddagger_j =R^\ddagger_{j-k}.$ Given $\delta$, the break-points are defined as $\hat{\tau}_f = \mbox{argmax}_{j \in {\cal M}_f}\{R_{j}^\ddagger: \, L_{j,4} \le \delta\}$ if ${\cal M}_f$ is the transition region from $W_0$ to $W_2$. If ${\cal M}_f$ is the transition region from $W_2$ to $W_0,$ $\hat{\tau}_f = \mbox{argmax}_{j \in {\cal M}_f}\{L_{j}^\ddagger: \, R_{j,4} \le \delta\}$.
\end{algorithm}

\begin{thm}
\label{th:multitwo-sided}
Assume Condition \ref{cond:two-sided-H1} and $Z_{i j}$ are $\sigma^2$-sub-Gaussian. Let $\gamma = c_1 (\log p)^{1/2}$, $\delta = c_2  (\log l)^{1/2}$, where $c_1, c_2 > 0$ are sufficiently large constants. Assume that $(\log p)^{1/2} = o(d^2 n \sqrt{k})$. Then there exists a constant $c > 0$ independent of $n$, $k$ and $p$ such that  
\begin{eqnarray}
\label{eq:06061632}
P\left[ \hat l = l, \, \max_{j \le l} |\hat \tau_j - \tau_j| \le {c k^{1/2} (\log l)^{1/2} \over n d^2} \right] \to 1.
\end{eqnarray}
\end{thm}


\section{Numerical studies}
\label{sec:03161012}
In this section, we present simulation studies, assess the finite sample performance of the proposed methods and compare them with competing methods \cite{fan07, yao1993, bh95}. We look at {one- and two-sided tests with one realization in Section \ref{sec:simu1} and Section \ref{sec:simu2}, respectively.} {One- and two-sided tests with more realizations} are presented in the Supplementary Material.

\subsection{Simulation study 1}
\label{sec:simu1}
Consider the model
$
X_i = \mu_i + Z_i, 1\le i \le p.
$
The number of tests are $p = 600, 2000$ and $6000.$ There are $2$ break-points $\tau_1 = 1+ 0.4p, \tau_2 = 0.6p,$ $1$ signal cluster $[\tau_1, \tau_2]$ and the configuration is displayed in Table 1 and Figure 3. We compare it with the change point detection for epidemic alternative proposed in \cite{yao1993}. We simulate data with three different error terms: standard normal distribution, rescaled student $t$ distribution with $6$ degree of freedom ($t(6)/1.5^{0.5}$) and rescaled Laplace distribution  ($\mbox{LP}(0, 1)/2^{0.5}$) so that their variances are all $1$.

The sliding window length $k = \lfloor p^{1/2}\rfloor$ is used in the calculation of  
$
R_i^{\circ} = k^{-1}\sum_{j=i+1}^{i+k}X_j.
$
We also show results for other choices of $k.$
In order to estimate the variance $\sigma^2$, we choose the tuning parameter $m = k.$ Let $p' = p-m+1$,
$
\hat{\sigma}_i^2 = m^{-1}\sum_{j=i}^{i+m-1}X_i^2, 1\le i \le p'.
$
Theoretically speaking, any statistics $\hat{\sigma}^2_{(j)}$ with $j \le p'/2$ are consistent and we use $\hat{\sigma}^2_{(\lfloor{p'/2 \rfloor})}$ as the estimate.

\begin{minipage}{\textwidth}
\noindent \begin{minipage}[b]{0.4 \textwidth}
\centering
\captionof{table}{Signal configuration for the one-sided test. seq 1: the linear sequence from 0.4 to 1.6; seq 2: the linear sequence from 1.6 to 0.4. Segment means percentage of the sequence.}
\begin{tabular}{lllll}
\hline
\noalign{\smallskip}
Segment  & 40 & 10 & 10 & 40\\
Signal & 0 & seq 1 & seq 2 & 0\\
\hline
\end{tabular}
\end{minipage}
\hfill
\begin{minipage}[t]{0.58\textwidth}
\centering
\includegraphics[width=0.58\textwidth]{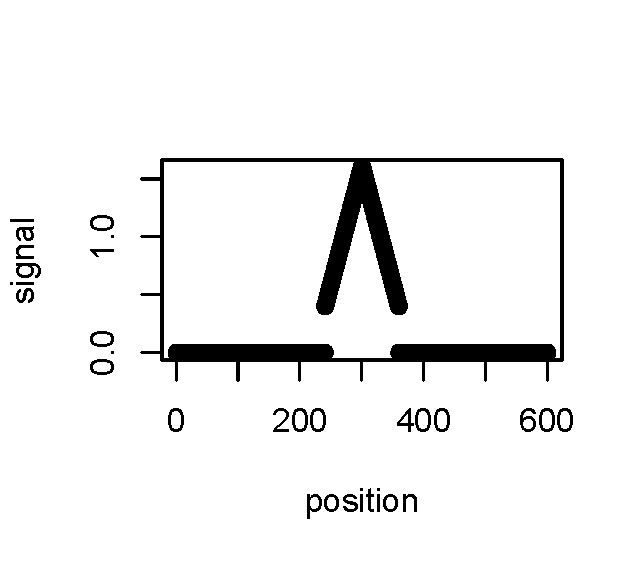}
\captionof{figure}{Signal configuration when $p=600.$}
\end{minipage}
\end{minipage}

\vspace{0.5cm}

We implemented algorithm \ref{algorithm:onesided}. Thresholding values $\gamma$ and $\delta$ are chosen as $0.95$th quantile of $\hat{\sigma}\mbox{max}_{0\le j \le p-k}G_j^{\circ}$ and and $\hat{\sigma}\mbox{max}_{j\in W_1}G_j^{\circ},$ respectively, where $G_j^{\circ} = \sum_{i=j+1}^{j+k}\eta_i/k, \eta_i, i \in \mathbb Z,$ are i.i.d. $N(0,1)$ and $W_1$ are the major connected components which include indices $j$ such that $Q_j^{\circ} = 1(R_j^{\circ} > \gamma) + 1(L_j^{\circ} > \gamma)=1.$  

In implementing \cite{yao1993}, we use {$L_1$,} the likelihood ratio statistic as an example for illustration. Similar results can be obtained for other test statistics. Specifically,
\begin{equation}\label{yao}
{L_1} = \mbox{max}_{1\le i<j\le p}\{\sum_{k=i+1}^{j}X_k - \frac{j-i}{p}\sum_{k=1}^{p}X_k - \frac{1}{2}\delta_0(j-i) \},
\end{equation}
where $\delta_0$ is the signal magnitude, which is assumed to be the same within a cluster in \cite{yao1993}. In our setup, we take $\delta_0 = 1,$ which is the average of signal magnitude within the cluster $[\tau_1, \tau_2].$ We identify the region $[\hat{I}, \hat{J}]$ as the epidemic alternative, where $\sum_{k=\hat{I}+1}^{\hat{J}}X_k - p^{-1}(\hat{J}-\hat{I})\sum_{k=1}^{p}X_k - \frac{1}{2}\delta_0(\hat{J}-\hat{I})$ is the obtained maximum value in (\ref{yao}). Note that the computational speed is quadratic with number of tests $p.$ Our evaluation criterion is the combined error rate (CER), which is the expected value of the ratio of the number of falsely rejected hypotheses and falsely accepted hypotheses over total number of  tests, the estimated number of break points $\hat{l}$ and the average difference between the estimated break points and true break points. 
For the proposed method, we also look at false discovery rate (FDR), which is the expected value of the ratio of false rejections over total rejections and the power, which is the expected value of the the ratio of true rejections over total number of non-nulls. 



Table \ref{t2} summarizes results based on $10^3$ replications. We can see that across different error distributions, the variance estimate $\hat{\sigma}^2$ has a decent performance and, as expected from our asymptotic theory, it is close to the true ones. The proposed method has smaller CER compared to method based on \cite{yao1993}, especially with large number of tests. Both procedures correctly identified 2 break points. The difference between estimated break points and true ones are smaller based on the proposed method especially with large samples. Our results are robust to different error terms and the sliding window length $k.$ For different error distributions the respective values of CER are quite close, as expected from our theoretical result.

\begin{table}[!htbp] 
\caption{Summary statistics for one-sided test with $1,000$ simulations.  $N(0, 1)$: standard normal; $t(6)/1.5^{0.5}$: rescaled student $t$ distribution with df $6$; $\mbox{LP}(0,1)/2^{0.5}$: rescaled Laplace distribution; 
$k$ is the window size; $\mbox{CER}$ is computed based on the proposed method; $\mbox{CER}_Y$ is based on \cite{yao1993}; $\hat{l}$ is estimated number of break points based on the proposed method; $\hat{l}_Y$ is estimated number of change points based on \cite{yao1993}; $\mbox{Diff}$ is the average distance between estimated break points and true break points based on the proposed method; $\mbox{Diff}_Y$ is the average distance between estimated change points and true change points based on \cite{yao1993}; {FDR is the expected value of the ratio of false rejections over total rejections and Power is the expected value of the the ratio of true rejections over total number of non-nulls.}   } 
\label{t2}
\begin{center}\footnotesize
\begin{tabular}{cccccccccc}
\hline
  $k$& $\hat{\sigma}^2$ & $\mbox{CER}$ &$\mbox{CER}_Y$ &  $\hat{l}$& $\hat{l}_Y$& $\mbox{Diff}$& $\mbox{Diff}_Y$ &FDR & Power\\
$p=600$ &&&&&&&&&\\
&&&&\multicolumn{2}{c}{$N(0,1)$} &&&&\\
24 & 1.0533 &0.0503  & 0.0538 &2  &2  &15.35  & 16.63 &  0.0016&0.75 \\
30 &  1.0665&0.0475  &0.0508  &2  &2  & 14.30 & 15.68 &0.0029 &0.77 \\
36 &  1.0605&0.0528  &0.0492  &2  &2  &15.85  &15.24  &0.0021  &0.74 \\
&&&&\multicolumn{2}{c}{$t(6)/1.5^{0.5}$} &&&&\\
24 & 1.0363 &0.0489  &0.0513  &2 &2 &14.66  &15.84   &0.0015  & 0.76\\
30 &  1.0312& 0.0511 &0.0533  &2 & 2 &15.33  &16.44   &0.0019  &0.75 \\
36 & 1.0425 &0.0554  &0.0543  &2 & 2 &16.61  &16.75  &0.0020  &0.73 \\
&&&&\multicolumn{2}{c}{$\mbox{LP}(0,1)/2^{0.5}$} &&&&\\
24 & 1.0128 &0.0517  &0.0528  &2  &2  &18.50  &16.33  &0.0033  & 0.74\\
30 &  1.0377&0.0532  &0.0548  &2  &2  &17.08  &16.85  &0.0051  &0.74 \\
36 & 1.0630 &0.0528  &0.0497  &2  &2  &15.84  &15.36  &0.0010  &0.74 \\
 $p=2,000$&&&&&&&&&\\
 &&&&\multicolumn{2}{c}{$N(0,1)$} &&&&\\
44 & 1.0469 &0.0262  &0.0495  &2  &2  &29.54  &50.05  &0.0021  &0.87 \\
55&  1.0535& 0.0244 &0.0514  &2  &2  &25.96  &51.90  &0.0016  &0.88 \\ 
66 & 1.0420 &0.0251  &0.0499  &2  &2 &25.15  &50.36  &0.0025 & 0.88\\
 &&&&\multicolumn{2}{c}{$t(6)/1.5^{0.5}$} &&&&\\
44 &1.0342  &0.0279  &0.0505  &2  &2  &31.46  &50.97  &0.0015  &0.86 \\
55&1.0394  &0.0248  &0.0503  &2  &2  &26.64  &50.82  &0.0019  & 0.88\\ 
66 &1.0355 &0.0265  &0.0518  &2  &2 &26.52  &52.28  &0.0032 &0.87 \\
&&&&\multicolumn{2}{c}{$\mbox{LP}(0,1)/2^{0.5}$} &&&&\\
44 & 1.0382 &0.0278  &0.0524  &2  &2  &32.84  &52.91  &0.0006  &0.86 \\
55& 1.0569& 0.0228 &0.0500  &2  &2  &22.77  &50.46  &0.0018  &0.89 \\ 
66 &1.0475 &0.0111  &0.0498  &2  &2 &26.15  & 50.33 &0.0017 & 0.87\\
$p=6,000$&&&&&&&&&\\
&&&&\multicolumn{2}{c}{$N(0,1)$} &&&&\\
60 &1.0433 &0.0170  &0.0495  &2  &2  &51.51  &148.85  &0.0007  &0.92 \\
77& 1.0396&0.0116  &0.0489  &2  &2  &40.67  &147.34  &0.0009  &0.94 \\
100& 1.0395 &0.0108 &0.0509 &2  &2 &32.34  &153.22    &0.0015  & 0.95\\
&&&&\multicolumn{2}{c}{$t(6)/1.5^{0.5}$} &&&&\\
60 &1.0308  &0.0168  &0.0505  &2  &2  &66.74  &152.02  &0.0008  &0.92 \\
77& 1.0330 &0.0127  &0.0505  &2  &2  &43.87  &151.99  &0.0014  &0.94 \\
100&  1.0345& 0.0103 &0.0507  &2 &2  &30.94    &152.71  &0.0011 &0.95\\
&&&&\multicolumn{2}{c}{$\mbox{LP}(0,1)/2^{0.5}$} &&&&\\
60 & 1.0429 &0.0181  &0.0493  &2  &2  &65.60  &148.27  &0.0004  &0.91 \\
77&  1.0465& 0.0136 &0.0499  &2  &2  &46.42  &150.07  &0.0009  &0.93 \\
100&1.0452  &0.0111 &0.0498 &2 &2 &33.16  &150.05    &0.0014  &0.95 \\
\hline
\end{tabular}\end{center}
\end{table}

{Per the request of a referee, we implement the BH procedure \cite{bh95} with Gaussian error term and summarize the results in Table \ref{tab:table}. The simulation set up is the same as that in Table \ref{t2}. At FDR level $5\%,$ we can see that the BH procedure always controls FDR but with low power for clustered signals.}

\begin{table}
\begin{center}
\caption{{BH procedure with Gaussian error term. The definition of FDR, Power and CER are the same as that in Table \ref{t2}}.}
\label{tab:table}
\begin{tabular}{c|c|c|c}
$p$  & $\mbox{FDR}_{\mbox{BH}}$ & $\mbox{Power}_{\mbox{BH}}$ & $\mbox{CER}_{\mbox{BH}}$\\
\hline
600& 0.0435 & 0.0011	&0.0724\\
6000&0.0433 & 0.0003 & 0.1101
\end{tabular}
\end{center}
\end{table}

{We also conduct simulation studies to check the empirical type-I error rates under the global null with Gaussian error term. The results are summarized in Table \ref{tab2}. At significance level $5\%,$ the proposed method has a similar type-I error rate to BH procedure under the global null as evidenced from Table \ref{tab2}.}

\begin{table}
\begin{center}
\caption{{Type-I error rates under the global null with Gaussian error term. $\mbox{FDR}$ represents FDR based on the proposed method and $\mbox{FDR}_{\mbox{BH}}$ represents FDR based on BH procedure. }}
\label{tab2}
\begin{tabular}{c|c|c|c}
$p$  & $k$& FDR  &$\mbox{FDR}_{\mbox{BH}}$ \\
\hline
600& 36 & 0.0594 & 0.0495 \\
6000&60 & 0.0396 & 0.0495
\end{tabular}
\end{center}
\end{table}

\subsection{Simulation study 2} 
\label{sec:simu2}
In this section, we examine the two-sided test procedure. Data is generated through model (2.1). Let $p = 600, 2000$ and $6000$. The signal configuration is summarized in Table \ref{t3}. We look at the robustness of our procedure with different error terms ($N(0, 1)/2^{0.25}$, $t(10)/(75/16)^{0.25}$ and $LP(0,1)/20^{0.25}$), which are standardized to have $\kappa=1$. Window size $k = \lfloor p^{1/2}\rfloor$ and $m = \lfloor p^{1/2} \rfloor$ are used for illustration. The calculation of $\hat{\sigma}^2, \gamma$ and $\delta$ are the same as that in simulation study 1 except that $\hat{\kappa}$ is used instead of $\hat{\sigma}$ and the calculation of $\hat{\kappa}$ is through $\hat{\kappa}^2 = \hat{\nu}_{(k)}/2 - 4\hat{\sigma}^4_{(k)}$. We follow Algorithm 2.2 to implement our method. As a comparison, results based on true values of $\sigma^2$ and $\kappa$ are presented as well.  
\begin{table}[htbp] 
\caption{Signal configuration for the two-sided test.  "-1 and 1 alternating": $\mu_i$ is $-1$ if $i$ is odd and $1$ if $i$ is even, $\mbox{seq}(0.5, 1.5)$: a linear sequence from $0.5$ to $1.5$ and $\mbox{seq}(1.5, 0.5)$: a linear sequence from $1.5$ to $0.5.$ }
\label{t3}
\begin{center}\footnotesize
\begin{tabular}{lcccccc}
\hline
Segment ($\%$ ) & 30 &10 &20 & 5 & 5 & 30\\
Signal strength  & 0 & $-1$ and $1$ alternating & 0 &$\mbox{seq}(0.5,1.5)$ & $\mbox{seq}(1.5, 0.5)$ & 0\\ 
\hline
\end{tabular}
\end{center}
\end{table}
%
%
From Table \ref{t4} we can see that procedures using the estimated parameters and the true ones have a comparable performance in terms of CER, FDR, power, estimated number of break points and the difference between estimated break points and true break points. This is consistent with our large sample theory. The results are relatively robust across different error terms. As numbers of tests increase, CER and FDR decrease and power and the difference between estimated break points and true break points increase. 

\begin{table}[htbp]
\caption{Summary statistics for two-sided test with $1,000$ simulations. Underscore $e$ is based on estimated $\sigma^2$ and $\kappa$, and underscore $t$ is based on true $\sigma^2$ and $\kappa$.}
 \label{t4}
\begin{center}\footnotesize
\begin{tabular}{ccccccccccc}
\hline
 p&$\mbox{CER}_e$ & $\mbox{CER}_t$ & $\mbox{FDR}_e$ & $\mbox{FDR}_t$ & $\mbox{Power}_e$ & $\mbox{Power}_t$& $\hat{l}_e$ & $\hat{l}_t$ & $\mbox{Diff}_e$ & $\mbox{Diff}_t$ \\
&&&&&&&&&&\\
$N(0,1)/2^{0.25}$&&&&&&&&&& \\
600 & 0.0822 & 0.0743 & 0.0207 & 0.0323 & 0.61 & 0.65 & 4 & 4 & 19.67 & 21.53\\
2000 & 0.0390 & 0.0354 & 0.0091 & 0.0142 & 0.81 & 0.84 & 4 & 4 & 19.47 & 19.81\\
6000&0.0223 & 0.0202 & 0.0048 & 0.0101 & 0.89 & 0.91 & 4 & 4 & 35.64 & 38.68\\
&&&&&&&&&&\\
$t(10)/(75/16)^{0.25}$&&&&&&&&&& \\
600 & 0.0817 & 0.0753 & 0.0266 & 0.0367 & 0.61 & 0.65 & 4 & 4 & 17.62 & 18.84\\
2000&0.0378 & 0.0339 & 0.0129 & 0.0203 & 0.82 & 0.85 & 4 & 4 & 36.37 & 30.40\\
6000& 0.0208 & 0.0192 & 0.0073 & 0.0109 & 0.90 & 0.91 & 4 & 4 & 51.26 & 64.54\\
&&&&&&&&&&\\
$LP(0,1)/20^{0.25}$&&&&&&&&&& \\
600& 0.0758 & 0.0642 & 0.0217 & 0.0295 & 0.64 & 0.71 & 4 & 4 & 18.53 & 13.03\\
2000&0.0350& 0.0327 & 0.0140 & 0.0173 & 0.84 & 0.85 & 4 & 4 & 35.70 & 41.91\\
6000&0.0199 & 0.0178 & 0.0078 & 0.0103 & 0.91 & 0.92 & 4 & 4 & 79.76 & 79.63\\
\end{tabular}
\end{center}
\end{table}

\section{Applications to real data}
\label{sec:real}
We now apply our procedure to an array-based Comparative Genomic Hybridization (array CGH) data. Array CGH is a powerful technology for measuring copy numbers at thousands of loci simultaneously. The output of array CGH experiment is usually a long vector, spanning each chromosome, recording the $\mbox{log}_2$ ratios of the normalized probe intensities from the test samples vs. the reference samples. These ratios of intensities are used to approximate the ratios of DNA copy numbers in the test samples vs. the reference samples. A $\mbox{log}_2$ ratio far from $0$ (either positive or negative) indicates a possible DNA copy number amplification or deletion for the probe. Identification of chromosomal alteration regions will provide valuable information to elucidate disease etiology and to discover novel disease related genes. 


In the study conducted by \cite{pollack02}, cDNA microarray CGH was profiled across $6691$ mapped human genes in $44$ breast tumor samples and $10$ breast cancer cell lines. The raw data can be downloaded from the PNAS website (\url{https://www.pnas.org/content/suppl/2002/09/23/162471999.DC1/4719CopyNoDatasetLegend.html}). We picked the breast cancer cell line BT474 as an example, and applied our method to detect DNA copy number amplification. Details of one realization are in the Supplementary Material and the results are presented in Table \ref{t7}

\begin{table}[!htbp]
\caption{Results based on one sequence and multiple sequence with one-sided test}
\label{t7}
\begin{center}\footnotesize
\begin{tabular}{cccccc}
\hline
\multicolumn{3}{c}{One realization} & \multicolumn{3}{c}{Multiple realizations}\\
Chromosome  & beginning  & ending &Chromosome  & beginning & ending \\
number & loci & loci & number & loci  & loci \\
11& 68434309& 81603744 & 11 & 46512342 & 81603744 \\
& & & 14 & 16522721 & 106822024\\
&&& 15&17156123&18891425\\
17& 28552955 & 82172608& 17&28552955 &42040770 \\
20&43585793&66314778&20& 44457372&66314778\\
21& 12430025 &15830914 & 21 & 12430025 & 15889676\\
\hline
\end{tabular}
\end{center}
\footnotesize{Note: Chromosome 14 and 15 are connected as one cluster with very short segments in chromosome 15 with multiple realizations analysis, chromosome 20 and 21 are connected as one cluster with both one realization and multiple realizations analysis.}
\end{table}

%

For multiple realization analysis, we consider the one-sided test using cell line 1 in addition to BT474 for analysis. We compute $\hat{\mu}_i$ and $\hat{\sigma}_i^2$ for $i = 1, \ldots, p$, and test statistics $\hat{R}_j = \sum_{l=j}^{j+k-1}\sqrt{2}\hat{\mu}_l/(\sum_{l=k}^{j+k-1}\hat{\sigma}_l^2)^{1/2}, j =1, \ldots, p-k+1.$ We use the same window length as in the one realization case $k = \lfloor p^{1/2} \rfloor = 78$, and compute ${Q}_j^{\star} = 1({R}_j^{\star} > \gamma) + 1(L_j^{\star} > \gamma)$ following algorithm \ref{algorithm:multionesided}. Critical values $\gamma = 3.8907$ and $\delta=1.0992$ are obtained through the $0.95$th quantile of the empirical distribution of $\mbox{max}_{1\le j \le p-k+1}{G}^{\star}_j$ and $\mbox{max}_{j \in W_1}{G}^{\star}_j,$ respectively, where $G_j^{\star} = \sum_{l=j}^{j+k-1}\hat{\sigma}_l\eta_l/(\sum_{l=j}^{j+k-1}\hat{\sigma}^2_l)^{1/2}, j \in \mathbb Z, \eta_j$  are i.i.d. $N(0,1)$ random variables and $W_1$ is the transition region which includes indices $j$ such that the smoothed $\tilde{Q}_j^{\star}=1$. 

The results are summarized in Table \ref{t7}. We can see that four clustered regions are detected by the multiple realizations analysis, three of which overlap with those detected by one realization analysis, 
which shows that amplifications in these genome regions are shared among the two breast cancer patients. The identified chromosomal amplification regions are implicated in the literature to harbor genes associated with breast cancer \citep{pollack99}. In cancer studies, ``passenger" mutations tend to occur more or less randomly throughout the genome, and ``driver" mutations tend to cluster and favor certain genome positions containing functionally relevant genes. An important goal in the analysis of tumor cell lines is to find the ``driver" mutations, which play a functional role in driving tumor progression \citep{stratton09}. Thus our analysis can suggest followup studies and intervention strategies. We choose to conduct our data analysis at the genome level, rather than at the chromosome level because genome scale analysis allows the detection of copy number aberrations involving entire chromosome arms, which might be missed in chromosome-level analyses for which no actual changepoints exist. 
  
\bigskip

{\bf Acknowledgments.} We are grateful to two referees for their many helpful comments. The research is partially supported by an NSF grant.

\end{document}